\documentclass[aps,prl,twocolumn,showpacs,superscriptaddress,groupedaddress,nofootinbib]{revtex4}
\usepackage{graphicx}  
\usepackage{dcolumn}   
\usepackage{bm}        
\usepackage{amssymb}   
\usepackage{slashed}
\usepackage{graphicx}				
\usepackage{amsmath}
\usepackage{mathtools}
\usepackage{tikz}
\usetikzlibrary{arrows,backgrounds}
\usepackage[all]{xy}
\usepackage{yfonts}
\begin{document}

\title{Quantization, Holography and the Universal Coefficient Theorem}
\author{Andrei T. Patrascu}
\address{University College London, Department of Physics and Astronomy, London, WC1E 6BT, UK}
\begin{abstract}
I present a method of quantization using cohomology groups extended via coefficient groups of different types. This is possible according to 
the Universal Coefficient Theorem (UCT).
I also show that by using this method new features of quantum field theory not visible in the previous treatments emerge. 
The main argument is that several constructions considered as absolute until now may appear as relative, depending on individual 
choices of group structures needed to probe a topology. The universal coefficient theorem also gives information about how these structures as 
measured by different choices of groups, relate to each other. This may result in the formulation of new dualities and a deeper understanding of the relation between quantum field theories
and gravity.
\end{abstract}
\maketitle
\section{Introduction}
The quantization of gravity is a major unsolved problem [1]. The equivalence principle [2], the black hole information paradox [3], the holographic conjecture [4], 
emergence of space-time [5] or coarse graining of observables [6] are only a few concepts that followed from it.
I present here a method that makes use of a theorem of algebraic topology and homological algebra (the universal coefficient theorem) in order to prove that some 
theoretical constructions used in previous descriptions of quantum gravity may not have an absolute meaning independent of some arbitrary choices of groups of coefficients.
These choices of coefficients may induce different topological structures, therefore assuming independence of coefficient groups implies a form of independence of 
topology. 
 The reason for considering this invariance
 as important in a quantum theory of gravity is the fact that there exist arbitrary choices that may make the connectivity of a space change.
One can cite the formation of a black hole that makes matter in a region of spacetime collapse onto itself. After the collapse passes the horizon there is no method of avoiding 
the central region where quantum effects like spacetime topology change may appear.  
Another example is the choice of making extremely accurate length measurements in space. This implies adding energy in a given 
region. This may in the end generate horizons which imply the collapse of matter towards a region where quantum gravity and changes of topology are assumed to be possible. 
One may assume therefore that a 
full theory of quantum gravity may not depend on arbitrary choices of this kind in the same way in which the formal aspects of general relativity should not depend on a choice of a 
coordinate system. 
The applicability of the theorem is not restricted to space-time itself but can be used generally to field-spaces, groups, various manifolds or discrete spaces. 
Its use in these different situations will be made implicitly. 
The main idea of this paper is that the identification of relevant physical observables in the QFT context is strongly dependent on the choice of coefficient groups
associated to (co)homology groups of the field space. 
The (co)homological structure of a field theory can be described with various coefficient groups, each inducing some indexation over the field space. It is well known 
that some choices are better than other. In general one uses a $\mathbb{Z}_{2}$-group when orientation is not relevant
or a $\mathbb{R}$-coefficient structure when continuum properties of the analyzed space appear to be relevant. 
However, there are more subtle applications of the coefficient groups.
I show here that the choice of one coefficient group instead of another can hide a set of physically relevant observables in the quantization procedure. Also, the 
logical assignment of observables in an equivalence class dictated by the availability of a practical measurement of its spectrum by an observer may allow, by using the 
axiom of choice, the construction of predictors for the spectrum of other observables in the same equivalence class [7].
As a result, it appears to be impossible to assign an absolute topology to a space (be it ``physical'' spacetime or the space of field configurations) in the 
absence of an arbitrary choice of a coefficient group.
I start with a field theoretical context. At this level already some aspects must be clarified. When quantizing a one particle theory one may use for example Feynman's path 
integral formulation. This implies the existence of an ``expectation catalogue'' for positions in space-time indexed in some way. As no information about the intermediate steps
is available one uses the principle of quantum mechanics that states that no actual state can be assigned to an object unless that state can be actually empirically confirmed
to be realized. In this case the integration that gives rise to the quantum amplitude must be a sum over all possible configurations. 
An extension of this principle was necessary due to the Lorentz group. As one was not able to discuss in the context of special relativity about a predefined or fixed number of
particles, quantum fields had to be introduced. These are simply extensions of the ``expectation catalogues'' of simple one-particle quantum mechanics. 
They are not ``measurable'' in any physical sense individually, but their interference and their topology is probed statistically by the rules of quantum mechanics. 
It should be well known that the statistics of an experiment (say Bohm-Aharonov) depends on the topology of the field space (the regions where the wavefunction is defined). 
In the end, the statistics must probe all connected components of all possible configurations. 
In the case of quantum gravity there are different approaches on how a quantum field theoretical formulation should look like. 
It is however clear that such a formulation should exist. I refer here to the works on string field theory [21].
 There the ``quantum field'' becomes a world-sheet-string-field
 ``expectation catalogue'' which is expanded even more with respect to the previous situations. 
While a string-field theoretical approach exists, it is not clear how the various configurations interrelate and what configurations can exist in various situations. 
Dualities are supposed to help in this aspect by identifying configurations and simplifying the overall problem. 
It appears to me that there exists a general method of constructing such dualities based on the ideas presented in this article. It also appears to me that 
the constructed dualities will have an applicability restricted to specific arbitrary choices of group-structures in topology. This is conjectured to be
 valid also for the holographic 
principle. It is the universal coefficient theorem 
that will in the end provide a description of what configurations can be simultaneously known and what configurations will interfere at the level of the ``catalogue of 
expectations''. It also appears that the change of topology is of major importance in quantum gravity as one expects a change in the topology of spacetime during the formation of
a black hole. However, the form of the laws of nature should not depend on a specific topology.
I partially follow in this introduction reference [8].

First construct a functor $\textfrak{E}$ from the category of spacetimes $(Loc)$ to the category of local convex vector spaces $(Vec)$.

This functor associates to each spacetime $M$ a configuration space $\textfrak{E}(M)$ of fields defined on it. 
The isometric embeddings $\chi : M\rightarrow N$ are mapped into 
pullbacks $\chi^{*}: \textfrak{E}(N)\rightarrow \textfrak{E}(M)$. 
The space of the observables called $\textfrak{F}$ will be the space of the functionals $F:\textfrak{E}(M)\rightarrow \mathbb{R}$. 
It is at this point that one also has to define the topological structure of the space (or space-time $M$). Physically this remains uncertain unless a choice of a 
coefficient group in (co)homology is made. This will define the topology and will allow a specific definition of the observables.
Essentially the ``experimental setup'' (or a coefficient group choice) tells spacetime how to connect. This connection tells quantum mechanics how the correlations between
``expectation catalogues'' should be constructed (what observables make physical sense). What follows is standard quantum mechanics which (via the universal coefficient theorem)
tells the experimentalist how to connect the results obtained with one group structure to possible results obtained by other observers using other group structures. 
This is important when one compares, for example, the observations made when falling towards a black hole to those of a far away observer.
Finally, accurate measurements and probing of spacetime at small scales implies adding energy in a small region of space which in the end may alter the topology of spacetime 
itself.

One can observe that in principle a topology induced by a choice of a coefficient group (via a particular experimental setup) results in a modified set of observables 
and a modified algebra for the resulting quantum (field) theory. Also, the geometry of the (field) space imposes restrictions on possible topologies (for example extreme 
curvature may imply restrictions over the allowed topologies). One can summarize this as 
\begin{widetext}

\[ Topology \left( \begin{array}{c}
\mbox{probed by quantum mechanics}\\
\mbox{induced by a choice of a coefficient group}
\end{array} 
\right)
\leftrightarrows
Geometry\left( \begin{array}{c}
\mbox{well defined local quantum observables} \\
 \mbox{quantum operator algebras}
\end{array} 
\right)\]
\end{widetext}
\par In this context the main question for quantum gravity is ``how do different geometries correlate?'' To this question one can give an answer when one considers the 
topology of the field space and the fact that this topology is not given in an absolute sense. The acceptance of the non-universality of topology (as proved clearly by 
the universal coefficient theorem) leads to different ``counting rules'' for different contexts.

In what follows one defines the class of functionals called ``local functionals'' as
\begin{equation}
 F(\phi)=\int_{M}dvol_{M}f(j_{x}(\phi))
\end{equation}
where $j_{x}(\phi)=(x,\phi(x),\partial\phi(x),...)$ is the jet of $\phi$ at the point $x$.

Let $L$ be a suitably defined Lagrangean. We can define an associated action functional $S[L[\phi]]$. The field equation becomes in this context $S'_{M}(\phi)=0$ where 
the prime denotes 
the Euler-Lagrange derivative. The space of solutions of this equation forms a subspace of $\textfrak{E}(M)$ called $\textfrak{E}_{S}(M)$. In the context of classical 
field theory
one is interested in the space of local functionals over $\textfrak{E}_{S}(M)$ called $\textfrak{F}_{S}(M)$. This space can be defined as the quotient 
$\textfrak{F}_{S}(M)$=$\textfrak{F}(M)$/$\textfrak{F}_{0}(M)$ where $\textfrak{F}_{0}(M)$ is the space of functionals that vanish on-shell (on $\textfrak{E}_{S}(M)$).
A (co)homological interpretation for the $\textfrak{F}_{S}(M)$ space is required. For this one needs a vector field structure on the configuration space. The action of the 
vector fields $X[.]$ on the space of smooth functionals $C^{\infty}(\textfrak{E}(M))$ is
\begin{equation}
 \partial_{X}F[\phi]=<F[\phi],X[\phi]>
\end{equation}
One can associate to the action functional a map from the set of test functions over the spacetime manifold to the space of 
``observable''-functionals $\delta_{S}:\textfrak{D}(M)\rightarrow\textfrak{F}(M)$ such that
\begin{equation}
\phi\mapsto <S_{M}'[\phi],X[\phi]>=\delta_{S}(X)(\phi)
\end{equation}
where $S_{M}'$ is the Euler-Lagrange derivative of the action. Suppose there is an action $S$ such that $\textfrak{F}_{0}(M)=\delta_{S}(\textfrak{D}(M))$. Then
\begin{equation}
 \textfrak{F}_{S}(M)=\textfrak{F}(M)/\textfrak{F}_{0}(M)=\textfrak{F}(M)/Im(\delta_{S})
\end{equation}
From this one can construct the chain complex
\begin{equation}
 0\rightarrow\textfrak{D}(M)\xrightarrow{\delta_{S}}\textfrak{F}(M)\rightarrow 0
\end{equation}
This can be associated with the Batalin-Vilkovisky complex used in the geometric quantization.
The 0-order homology of this complex is $\textfrak{F}_{S}(M)=\textfrak{F}(M)/\textfrak{F}_{0}(M)$.
The set of critical points of the action functional
\begin{equation}
\{\phi \in \textfrak{D}(M) | \delta_{S}[\phi]=0\}
\end{equation}
 contains connected components that can be identified by the first homotopy group
\begin{equation}
\pi_{0}(\{\phi \in \textfrak{D}(M) | \delta_{S}[\phi]=0\})
\end{equation} 
The functionals on the classes of this group are the gauge invariant observables. One can see that the correct
identification of possible maps as well as homotopically
equivalent structures is extremely important for the correct
construction of the field space in the phase preceding
actual quantization.
Probably the best mathematical formalization of quantum mechanics is offered by  what is known as ``geometric quantization'' [9]. In this formulation one starts with a 
classical theory and follows a set of steps that assure the consistency of the resulting quantum theory. One may start with a general
 classical action depending on a set of fields $S[\phi]$. This implies the existence of a symplectic manifold. The main idea is to realize the symplectic form of this
 manifold as the curvature of a $U(1)$ principal bundle with a connection. We obtain the pre-quantum Hilbert space as the Hilbert space of square integrable sections
 of the principal line bundle. One has to pick for each point in this space
 a certain subspace of the complexified tangent space at that point. One defines the quantum Hilbert space to be the space of all square integrable sections of the line
 bundle that give $0$ when differentiated covariantly at that point in the direction of any vector of the tangent space. 
As basic quantum mechanics teaches us there exist two sets of variables that become non-commutative operators when quantizing. These may be called ``positions'' 
and ``momenta'' although their physical meaning may be rather different. 
The next step is the choice of a polarization i.e. the choice of ``positions'' and ``momenta''. This choice is not unique. Once a polarization is available one 
can form a Hilbert space of states as the space of sections 
of the associated line bundle. The last step would be to associate to the classical variables actual quantum operators on the quantum Hilbert space. This amounts 
to the quantization of observables while mapping Poisson brackets to commutators. This procedure is in general not well defined for all operators. 
Strictly speaking the method of geometric quantization is not properly defined in the context of quantum gravity. The definition of a field-space or a space of
configurations is extremely complicated and the integration over such a structure appears to be ill-defined. However, it is precisely the method presented in this article that
may add some extra structure to this space (for example via the addition of new dualities) such that its rigorous definition might become possible.
Several attempts of using geometric quantization in the context of string theory are known [23],[24] but the subject remains open for future research. 
\par Given a BV complex and some quantum observables in the context of a choice of a coefficient structure I now state the following Lemma
\par
{\bf Lemma 1} (The Universal Coefficient Theorem)
\par If $C$ is a chain complex of free abelian groups, then there are natural short exact sequences 
\begin{equation}
 0\rightarrow H_{n}(C)\otimes G\rightarrow H_{n}(C;G)\rightarrow Tor(H_{n-1}(C),G)\rightarrow 0
\end{equation}
$\forall$ $n$, $G$, and these sequences split. Here  $Tor(H_{n-1}(C),G)$ is the torsion group associated to the homology.
In this way homology with arbitrary coefficients can be described in terms of homology with the ``universal'' coefficient group $\mathbb{Z}$
$\flat$
\par This lemma is also valid for cohomology groups where it is formulated as
\begin{widetext}
\begin{equation}
 0\rightarrow Ext(H_{i-1}(C),G)\rightarrow H^{i}(C;\mathbb{Z})\otimes G\xrightarrow{h}H^{i}(C;G)\xrightarrow{r}Hom(H_{i}(C),G)\rightarrow 0
\end{equation}
\end{widetext}
where now the $Tor$ group on the right is replaced by the $Ext$ group on the left.
Moreover, this theorem is a property of algebraic topology independent of the existence of an underlying manifold structure for 
the spaces or groups on which it may be applied. 
For a proof in both the homology and the cohomology cases see reference [10].
The following example shows how the choice of the coefficient group can affect the correct identification of the homotopy type of a function.
\par
{\bf Example 2} (Homotopy and coefficient group)
 \par Take a Moore space $M(\mathbb{Z}_{m},n)$ obtained from $S^{n}$ by attaching a cell $e^{n+1}$ by a map of degree $m$. 
The quotient map $f:X\rightarrow X/S^{n}=S^{n+1}$ induces trivial homomorphisms on the reduced homology with $\mathbb{Z}$ coefficients since the nonzero 
reduced homology groups of $X$ and $S^{n+1}$ occur in different dimensions. But with $\mathbb{Z}_{m}$ coefficients the situation changes, as we can see considering 
the long exact sequence of the pair $(X,S^{n})$, which contains the segment
\begin{equation}
 0=\tilde{H}_{n+1}(S^{n};\mathbb{Z}_{m})\rightarrow \tilde{H}_{n+1}(X;\mathbb{Z}_{m})\xrightarrow{f_{*}}\tilde{H}_{n+1}(X/S^{n};\mathbb{Z}_{m})
\end{equation}
Exactness requires that $f_{*}$ is injective, hence non-zero since $\tilde{H}_{n+1}(X;\mathbb{Z}_{m})$ is $\mathbb{Z}_{m}$, the cellular boundary map 
\begin{equation}
H_{n+1}(X^{n+1},X^{n};\mathbb{Z}_{m})\rightarrow H_{n}(X^{n},X^{n-1};\mathbb{Z}_{m})
\end{equation}
being exactly 
\begin{equation}
\mathbb{Z}_{m}\xrightarrow{m}\mathbb{Z}_{m}
\end{equation}
One can see that a map $f:X\rightarrow Y$ can have induced maps $f_{*}$ that are trivial for homology with $\mathbb{Z}$ coefficients but not so for 
homology with $\mathbb{Z}_{m}$ coefficients for suitably chosen $m$. This means that homology with $\mathbb{Z}_{m}$ coefficients can tell us that $f$ is not
 homotopic to a constant map, information that would remain invisible if one used only $\mathbb{Z}$-coefficients. 
$\flat$
\par 
As the final step of this introduction I state here the main theorems of this article as well as a conjecture.
\par
{\bf Theorem 1} (Relativity of Observables)
There exist observables visible using some choices of coefficient groups and invisible using other choices. $\flat$
\par
{\bf Theorem 2} (Relativity of distinguishability)
\par
There exists no unequivocal measure of distinguishability of quantum states that is independent of the choice of the coefficient group. Distinguishability is relative. $\flat$
\par
{\bf Theorem 3} (Relativity of Symmetry)
\par
A particular choice of a coefficient group makes a specific symmetry structure in the field space manifest. There exists no absolute symmetry. $\flat$
\par
{\bf Conjecture} (Relativity of Holography)
\par
There is no general unequivocal mapping of any consistent geometric structure in a space-time volume to its surface. In the full context of quantum gravity the 
existence of a holographic principle is an undecidable statement depending on particular choices of the coefficient groups. ``Strong-weak'' dualities can however be 
constructed and generalized in a case-by-case way $\flat$
\par
The proofs of the theorems as well as validity arguments for the conjecture are provided in the following chapters. 
The method of proof is as follows: 
I make a choice of a coefficient group in cohomology (i.e. a choice of topology).
I try to construct standard quantum mechanics (eventually using geometric quantization).
If geometric quantization is impossible I can always switch to a different topology where this method is possible and see how it relates 
to the topology where it was impossible via the universal coefficient theorem. This may bring new insights about the geometric quantization prescription.
I construct a set of observables and physical states using a particular choice of the coefficient group. 
I obtain a set of physical states obeying some properties (distinguishability, etc.). 
I make another choice of the group structure where the above stated properties are not valid any more. By the Universal Coefficient Theorem it follows that the considered 
properties are relative i.e. cannot be associated to a full theory of quantum gravity. 
\section {Path integral quantization and field theories}

\par 
One method of quantization is given by what is known under the name of ``Feynman path integral'' [22]. 
This has been generalized although not completely, for string field theory [21].
For an introduction I partly follow [22]. 
I assume that the standard prescription of computing quantum probabilities using 
quantum amplitudes is well known. 
If $P_{ac}$ is the quantum probability of measuring event $c$ when it follows the measurement of event $a$ then the probability must be
calculated as $P_{ac}=|\varphi_{ac}|^{2}$ where $\varphi_{ac}=\sum_{b}\varphi_{ab}\varphi_{bc}$ where the sum is over the possible intermediate states $b$ which, I emphasize, 
following Feynman (ref. [22], page 3 in manuscript) have no meaningful independent value. 
In a 1-space and 1-time dimensional context a succession of measurements may represent a succession of the space-coordinate $x$ at successive times $t_{1},t_{2},...$, 
where $t_{i+1}=t_{i}+\epsilon$. Let the observed value at $t_{i}$ be $x_{i}$. Classically the successive values
of $x_{1},x_{2},...$ define a path $x(t)$ when $\epsilon\rightarrow 0$. 
If the intermediate positions are actually measured one may talk about such a path with a well defined set of observed positions $x_{1},x_{2},...$ and the probability that 
the specified path $P(...x_{i},x_{i+1},...)$ lies in a region $R$ is given by the classical formula
\begin{equation}
 P=\int_{R}P(...x_{i},x_{i+1},...)...dx_{i}dx_{i+1}...
\end{equation}
where the integral is taken over the ranges of the variables which lie within the region $R$. 
If the intermediate positions are not measured then one cannot assign a value to them. In this case the probability of finding the outcome of a measurement in $R$ is
$|\varphi(R)|^{2}$ and $\varphi(R)$, i.e. the probability amplitude, is calculated as
\begin{equation}
 \varphi(R)=\lim_{\epsilon\rightarrow 0}\int_{R}\Phi(...x_{i},x_{i+1},...)
\end{equation}
where $\Phi(...x_{i},x_{i+1},...)$ defines the path. In the given limit this object becomes a path functional. 
There should be no mystery nowadays that the probability amplitude should be calculated as 

\begin{equation}
 \varphi(R)=\lim_{\epsilon\rightarrow\ 0}\int_{R}exp[\frac{i}{\hbar}\sum_{i}S(x_{i+1},x_{i})]...\frac{dx_{i+1}}{A}\frac{dx_{i}}{A}...
\end{equation}
where $S$ is the action functional for the given path segment. 
In order to go a step further and define the wavefunction in this context I will continue to follow Feynman's paper [22]. The region $R$ considered above can be divided
into future and past with respect to a choice of a time position $t$. One can define the region $R'$ as the past and the region $R''$ as the future. 
The probability amplitude connecting these regions will be
\begin{equation}
 \varphi(R',R'')=\int \chi^{*}(x,t)\psi(x,t)dx
\end{equation}
where
\begin{equation}
 \psi(x_{k},t)=\lim_{\epsilon\rightarrow 0} \int_{R'}exp[\frac{i}{\hbar}\sum_{i=-\infty}^{k-1}S(x_{i+1},x_{i})]\frac{dx_{k-1}}{A}\frac{dx_{k-2}}{A}...
\end{equation}
and
\begin{equation}
 \chi^{*}(x_{k},t)=\lim_{\epsilon\rightarrow 0} \int_{R''}exp[\frac{i}{\hbar}\sum_{i=k}^{\infty}S(x_{i+1},x_{i})]\frac{1}{A}\frac{dx_{k+1}}{A}\frac{dx_{k+2}}{A}...
\end{equation}
In this way one can separate the ``past'' and the ``future'' via the functions $\psi$ and $\chi$.
One may also construct a closer equivalence to the matrix representation of quantum mechanics by introducing matrix elements of the form 
\begin{widetext}
\begin{equation}
<\chi_{t''}|F|\psi_{t'}>_{S}=\lim_{\epsilon\rightarrow\ 0}\int ... \int \chi^{*}(x'',t'')F(x_{0},...x_{j})exp[\frac{i}{\hbar}\sum_{i=0}^{j-1}S(x_{i+1},x_{i})]\psi(x',t')\frac{dx_{0}}{A}...\frac{dx_{j-1}}{A}dx_{j}
\end{equation}
\end{widetext}
In the limit $\epsilon\rightarrow 0$, $F$ is a functional of the path $x(t)$. 
At this moment one can define various equivalences between functionals. These are to be associated to operator equations in the matrix formulation. 
One can of course define $\frac{\partial F}{\partial x_{k}}$ and one can calculate the associated matrix element using an action functional $S$. 
Using the fact that the action functional appears as $exp(\frac{i}{\hbar} S)$ one obtains matrix equations as, say
\begin{equation}
 <\chi_{t''}|\frac{\partial F }{\partial x_{k}}|\psi_{t'}>_{S}=-\frac{i}{\hbar}<\chi_{t''}|F\frac{\partial S}{\partial x_{k}}|\psi_{t'}>_{S}
\end{equation}
which can be stated as a functional relation defined for an action $S$ as
\begin{equation}
 \frac{\partial F }{\partial x_{k}} \leftrightarrow -\frac{i}{\hbar}F\frac{\partial S}{\partial x_{k}}
\end{equation}
Using the fact that $S=\sum_{i=0}^{j-1}S(x_{i+1},x_{i})$ one can rewrite
\begin{equation}
  \frac{\partial F }{\partial x_{k}} \leftrightarrow -\frac{i}{\hbar}F[\frac{\partial S(x_{k+1},x_{k})}{\partial x_{k}}+\frac{\partial S(x_{k},x_{k-1})}{\partial x_{k}}]
\end{equation}
In the case of a simple 1-dimensional problem one can write 
\begin{equation}
\frac{\partial S(x_{k+1},x_{k})}{\partial x_{k}}=-m(x_{k+1}-x_{k})/\epsilon
\end{equation}
and
\begin{equation}
 \frac{\partial S(x_{k},x_{k-1})}{\partial x_{k}}=+m(x_{k}-x_{k-1})/\epsilon-\epsilon V'(x_{k})
\end{equation}
Neglecting terms of order $\epsilon$ one obtains
\begin{equation}
 m\frac{(x_{k+1}-x_{k})}{\epsilon}x_{k}-m\frac{(x_{k}-x_{k-1})}{\epsilon}x_{k} \leftrightarrow \frac{\hbar}{i}
\end{equation}
The important aspect here is that the order of terms in a matrix operator product corresponds to the order in ``time'' of the corresponding factors in a functional. 
The order of the factors in the functional is of no importance as long as the indexation of these factors is reflected in the ordering of the operators in the
matrix representation. This means the left-most term in the above equation must change order so that one obtains the well known commutation relation
\begin{equation}
 px-xp=\frac{\hbar}{i}
\end{equation}
\par
One may observe that the choice of a specific indexation of the measurement outcomes, according to a time index 
(i.e. $\mathbb{Z}$-group), leads to the well known commutation relations. 
The ideas behind path integral quantization are kept intact when going to the relativistic context. However, when we have to go to a gravitational context the sum over 
configurations (geometries)
becomes non-trivial. In this sense one has to construct the (co)homology structure of the space and one has to deal with the universal coefficient theorem. 

This theorem states that 
a specific framework, constructed through the choice of a coefficient group in (co)homology is, up to (extension) torsion in (co)homology, equivalent with the choice of an integer 
coefficient group. However, some choices of coefficient groups may make some observables manifest while others may hide them. Moreover, simple order relations as the ones 
used in the proof above are no longer uniquely defined. What was identified by Feynman as a natural choice (time ordering) may in fact be just the result of 
a given coefficient group. Other ordering relations (like radial ordering in the case of CFT's) are also known.
It is visible in this context that the construction of a path integral prescription using another coefficient group will change the quantization prescription (as formulated via 
the algebra of operators). Quantization doesn't mean only algebra of operators, as has been made obvious in the definition of geometric quantization. 
In an ideal situation one would expect a physical motivation that determines the operator algebra. This might appear in the context of the application of universal coefficient
theorems. 
The group structure imposed over the configuration space can be chosen for example as $\mathbb{R}/\mathbb{Z}$ case in which one arrives at a continuous cyclic structure. 
This will present a somehow altered operator algebra. 
One may ask what is the physical meaning of the coefficient group? 
In fact, it is an extra layer of information that has to be dealt
with when performing quantization. It appears that it is not sufficient to simply integrate over non-equivalent field configurations as done in non-gravitational models. The
coefficient structure adds new ``degrees of freedom'' to the problem. These must be considered when performing path integral quantization in order to obtain suitable unitary
results. From this point of view, the extra-structure appears to be a step forward towards the unambiguous solution of the unitarity problem 
(also known as ``information paradox''). In a less formal tone, the ``information'' describing the system is encoded not only in the actual system but also in the 
set of rules one chooses in order to ``read'' that information.
I stressed in the above digression that the intermediate states in the path integral formulation must be added to the amplitude while 
keeping all possible outcomes, mainly because one cannot assign an outcome before a measurement is performed. The same considerations are valid when dealing with coefficient 
groups. While one can certainly prepare an experiment that involves a special choice of a coefficient group one will obtain a result dependent of this choice. When no 
practical choice is made one cannot assign any ``physical'' value to the choices of coefficients but one must consider them when calculating quantum amplitudes. 
From this perspective the question of the existence of a ``Planck scale topology'' is void of meaning. ``Microscopic geometries'' are to be associated to choices of coefficient
groups and these choices are arbitrary.
However, the universal coefficient theorem generates classes of topologies that can be identified in the sense of having the same $Ext$ and $Tor$ groups.
This may lead to an overall simplification of the path integral formulation as many configurations will appear as connected by dualities.  
One should notice that both string theory and quantum loop gravity assume special choices of topology as being absolute (Lie group topology for string theory as the ``string 
worldsheet'' and discrete topology for LQG). I consider these choices as an epistemological issue. 
In string theory one starts by postulating a fundamental string. This implies a continuous group structure and a well defined topology. By the universal coefficient theorem
however, this is simply a convention. Using that convention one arrives at an algebra of operators (say, Virasoro algebra). It should be clear now that this choice has nothing
fundamental to it. 
In quantum loop gravity one starts the other way around: one fixes the canonical quantization prescription involving the standard algebra and obtains in the end 
a particular topology (a discrete topology). Again, one arbitrary choice determines the other. There is nothing fundamental to it either. 
One cannot assign a precise topology to any space unless one makes a choice of 
a coefficient group in cohomology. In order to do this one must consider the universal coefficient theorem and its $Tor$ and $Ext$ groups. Any fixation on an absolute
topology would be equivalent with the postulation of the ``ether'' in special relativity i.e. void of meaning. 

One may notice that quantum gravity cannot be defined using a fixed (non-dynamical) spacetime manifold.
In fact, analysis in terms of the universal coefficient theorem makes the spacetime highly dynamical allowing even changes of topology. 
These can be seen if one considers for example coefficient groups of finite torsion degrees. 
The larger (but finite) the torsion degree of the group the more ``non-local'' will the associated ``observables'' look. 
The "non-local" behaviour in extreme conditions (black holes) is essentially the result of
a specific choice of topology. This will persist until clearer information about the group structure imposed by a particular experiment is given. 
When this happens is for the experiment to decide. The situation is similar to the supposed "objective collapse of the wavefunction" which is assumed 
(wrongly) to actually happen at some scale. This mistake vanishes when one understand that the wavefunction is to be interpreted as a "expectation catalogue". 
In the same way, when information about the connectivity of spacetime and of the "field-space" becomes manifest one will have to adopt the local structure at hand.
Of course, topologically disconnected macroscopic black holes may retain (from the perspective
of an observer lying outside) some apparent non-local aspects as their internal structure is inaccessible. 

One may ask if my method has as result the identification of different representations for the same algebra of operators. This is not the case. As can be seen from Feynman's example
the specific ordering of the events generates some commutation relations which define the algebra of operators. If one generalizes this to 
different choices of coefficient groups for probing the field space one can see that the algebra of operators will not be preserved. Indeed, one can use coefficients 
in a continuous group. In this case one can recover the string-theoretical case where a continuous line-like object appears as ``fundamental'' and in fact the algebra of its operators
is rather different. The associated group is generally not easily connected to the local algebra as the $Exp$ map is not always easily defined.
 Continuous group coefficients are useful. It is well known that one uses continuous coefficient groups when one wishes to avoid unnecessary complications 
due to the low-scale behaviour of the space to be studied. In fact, a claimed advantage of working with string-like objects is its so called ``UV-completeness''. Of course, from 
the perspective of coefficient-group-extended quantization this property is just a trade-off between using continuous groups in order to have UV-completeness and the
 complications that appear in the BRST-cohomology treatment of string theory.

\section{Relativity of Observables}
As shown in the introduction, the physical observables are to be identified with the functionals over the classes of the homotopy group associated to the critical points of the
action functional. Example 2 already showed how this identification is relativized by the UCT. I give here a more detailed proof. Take a set of observables obtained after
geometric quantization
\begin{equation}
 \mathcal{A}=\{A_{1},A_{2},...,A_{n}\}
\end{equation}
where $\mathcal{A}\subset\textfrak{F}_{S}$. While in the classical case $\textfrak{F}_{S}$ is to be associated with a space of local functionals, in the case of quantum gravity 
the locality condition may be relaxed (see ref. [11]). 
One can observe that the BV-complex 
\begin{equation}
0\rightarrow\textfrak{D}(M)\xrightarrow{\iota}\textfrak{F}(M)\xrightarrow{\gamma}\textfrak{F}_{S}(M)\rightarrow 0
\end{equation}
with $\textfrak{F}_{S}(M)=\textfrak{F}(M)/\textfrak{F}_{0}(M)$ and $\delta_{S}=\gamma \circ \iota$
can be represented as the complex of example 2
\begin{equation}
 0\rightarrow \tilde{H}_{n+1}(X;\mathbb{Z}_{m})\xrightarrow{f_{*}}\tilde{H}_{n+1}(X/S^{n};\mathbb{Z}_{m})\rightarrow ...
\end{equation}
In the last case $f_{*}$ is the induced map over the homology groups of the map $f:X\mapsto X/S^{n}$ over the analyzed spaces. In the case of the BV-complex the 
original maps would be the 
functionals $F:\textfrak{E}_{S}\mapsto \textfrak{E}_{S}$ which are to be associated to the physical observables of the quantum theory. 
In the same way as in example 2 one can define the map as a function of degree $m$. 
One may remark that observables that cannot be distinguished in $\mathbb{Z}$ will be visible if the choice of coefficients is $\mathbb{Z}_{m}$.

In order to have a correct representation of the actual set of observables one must redefine $\mathcal{A}$ as 
\begin{equation}
 \mathcal{\tilde{A}}=\{[A_{1}],[A_{2}],...,[A_{n}]\}
\end{equation}
where each term $[A_{i}]$ may be a set of observables on its own, the elements of which may not be discernible given a specific choice of coefficients.
It has been noted in reference [11] that for example classes of microscopical observables of black holes may be inaccessible to independent measurement
 due to large energies or long times required for 
accurate probing. While this is certainly possible I show here that the same can happen due to certain choices of coefficient groups. While it is certainly 
always possible to change the coefficient group with which one probes the field space this change may involve a change in the physical experimental setup. This would make
a simultaneous use of two coefficient groups in the same experiment impossible. As indiscernability of observables (coarse graining) may imply emergent locality 
(as shown in [11]) it may look like the UCT assures some form of locality at all levels. However, I am cautious in calling this ``locality'' with its proper name. I am also
 cautious when speaking about ``emergent locality'' or even more drastically, ``emergence of space-time'' (see ref. [5])
The reasons for this caution are expressed in the following section.

\section{Relativity of distinguishability}
Ongoing research in quantum information has led to various alternative definitions of distinguishability of quantum states. One recent paper [11] argues that physical
 criteria 
like extreme energy requirements or long waiting times would make some distinctions between quantum states impractical. I show here that in fact distinguishability of quantum
 states is mainly related to choices of the coefficient groups of (co)homology. There exist possible predictors that allow ``guesses'' concerning the presence of 
different physical states in the same equivalence classes associated to some observers [7].
Using quantum information tools one observes that given a set of observables $\mathcal{A}$ one cannot distinguish a random pure micro-state in a microcanonical ensemble $H_{E}$
of dimension $d_{E}$ from the maximally entangled state $\Omega_{E}=\frac{I_{E}}{d_{E}}$ unless the number of different outcomes of the operator $N(\mathcal{A})$ scales as
 $\sqrt{d_{E}}$.
Whenever $N(\mathcal{A})\sim \sqrt{d_{E}}$ one would require a long time or very large energies to achieve the accuracy that would allow the distinction of these states.
These statements presented also in [11] are partially correct. While one can follow the standard path of constructing normed or semi-normed spaces that 
would predict how ``far away'' quantum states
are in a given configuration I show here that these measures must be relative considering the fact that the arbitrary choice of a coefficient group may make the 
difference 
between distinguishability and indistinguishability of two quantum states relative. This statement is in full agreement with the uncertainty principle and
in the spirit of quantum mechanics as it extends the concept of uncertainty to the arbitrary choice of a coefficient group. 
In this section I follow ref. [11] in order to introduce the concepts I require. 
Consider a finite dimensional subspace $H_{E}\subset H$ of dimension $d_{E}$ consisting of all pure states $\psi=|\psi><\psi|$ that live in a microcanonical ensemble of energy
$[E-\delta E, E+\delta E]$. I may assume that the Hamiltonian describing the unitary time evolution of the system has non-degenerate energy gaps.
Consider again the set of observables $\mathcal{A}=\{A_{1},A_{2},...,A_{n}\}$. One may ask what are the necessary conditions for such a set to distinguish a random pure state 
$\psi \in H_{E}$ from a maximally mixed state in $H_{E}$.
One can follow two obvious paths and one less obvious path to quantify the difference between quantum states $\psi\in H_{E}$. 
What one obviously could do is to measure the expectation value of some operator $A\in\mathcal{A}$. However, the measurement of expectation values of an 
observable is not sensitive
enough to distinguish any different quantum states.
A quantum measurement in general offers a set of eigenvalues $a$ appearing with some probabilities $p_{a}$. Most of the information about the quantum system is encoded in the
probability spectrum $\{p_{a}\}$. Hence in order to distinguish two quantum states $\rho$ and $\sigma$ using a particular observable $A$ one can define a measure as
\begin{equation}
 D_{A}(\rho,\sigma)=\frac{1}{2}\sum_{a}|tr(|a><a|\rho)-tr(|a><a|\sigma)|
\end{equation}
$|a>$ being the eigenvectors of $A$. This measure is defined so that it encodes the information of the entire spectrum $\{p_{a}\}$. One can extremize the definition 
in order to define
a measure over a whole set of observables
\begin{equation}
 D_{\mathcal{A}}(\rho,\sigma)=\max_{A\in\mathcal{A}}D_{A}(\rho,\sigma)
\end{equation}
If $\mathcal{A}$ includes the entire set of observables in the Hilbert space one may define the distinguishability of two quantum states in general as
\begin{equation}
 D(\rho,\sigma)=\frac{1}{2}tr|\rho - \sigma|_{\mathcal{A}}
\end{equation}
where $|\rho - \sigma|_{\mathcal{A}}$ is the maximal difference in probability spectra over the entire set of available observables. 
If I continue to use this language it will be impossible to identify the restrictions due to the universal coefficient theorem. In fact one has to go a step
 back and to remember
that quantisation implies summation over inequivalent field configurations and this implies the construction of (co)homology groups. Physical observables are identified with 
the functionals over the classes of these groups. Different choices of coefficient groups in the (co)homology may lead to identification of functionals (they may appear as
homotopic to the identity) while using other groups may make them appear in different classes (i.e. being different observables). 
Considering that special features of the field space induced by mappings of finite degree cannot be ignored in the procedure of quantization one may have for a complex like
\begin{equation}
 0\rightarrow \tilde{H}_{n+1}(X;\mathbb{Z}_{m})\xrightarrow{f_{*}}\tilde{H}_{n+1}(X/S^{n};\mathbb{Z}_{m})\rightarrow ...
\end{equation}
a set of observables $\mathcal{A}=\{A_{1},A_{2},...,A_{n}\}$ while under
\begin{equation}
 0\rightarrow \tilde{H}_{n+1}(X;\mathbb{Z})\xrightarrow{f_{*}}\tilde{H}_{n+1}(X/S^{n};\mathbb{Z})\rightarrow ...
\end{equation}
another set $\mathcal{\tilde{A}}=\{[A_{1}...A_{i_{1}}],[A_{i_{2}}...A_{i_{3}}]...,[A_{i_{k}}...A_{i_{n}}]\}$
where the observables in the square brackets represent the classes of observables that cannot be distinguished in the given coefficient setup.
One may imagine that the choice of a coefficient group induces a forgetful functor between the category of observables $\mathcal{A}$ and $\mathcal{\tilde{A}}$. 
This functor also maps the discernability measure from  
\begin{equation}
D_{\mathcal{A}}(\rho,\sigma)=\max_{A\in\mathcal{A}}D_{A}(\rho,\sigma) 
\end{equation}
towards
\begin{equation}
D_{\mathcal{\tilde{A}}}(\rho,\sigma)=\max_{A\in\mathcal{\tilde{A}}}D_{A}(\rho,\sigma)
\end{equation}
One may observe that although the definition is still valid, the set of available observables changed significantly. One may look at this as a change of topological basis 
although this analysis may be beyond the scope of this article. 
In the last section I invited to caution in using terms like locality in relation to indiscernability of observables and entanglement. Indeed, the prescription of maximization
used in the definition of the measure above is not trivial. Following the universal coefficient theorem, in order to establish the maximum over the set of observables, one will 
always have to pick one element from an equivalence class. One may not be aware of the existence of more than one element in the given class but the class exists and
 a choice has
to be made in order to be able to compare in the end representatives from various classes. In order to be able to do this (as the elements of one class are supposed to be 
indiscernable so one cannot define a choice function) one has to invoke the axiom of choice. 
However, associating probability theory and the axiom of choice in the context of quantum mechanics is probably the most non-trivial task in mathematical logics.
Examples of how the axiom of choice reflects on the mathematics of coordinated inference can be found in [7]. A suitable analysis of these problems in the realm of
quantum information is the subject of a future paper. 
What I may add here is that the indexation of operators in $\mathcal{A}$ and $\mathcal{\tilde{A}}$ may give an order relation in terms of, for example, energy.
In this sense one may define the order over the operators in $\mathcal{A}$ as
\begin{equation}
 A_{1}\prec A_{2}\prec ... \prec A_{n}
\end{equation}
This ordering implies the visibility at a given energy. However, the deformation of some observables such that they enter a single homotopy class after the application of 
a new coefficient group may alter this order. In fact, one will have to define an order relation between equivalence classes where the choice of representatives is not 
unambiguously defined in the absence of the axiom of choice
\begin{equation}
[A_{i_{1}}]\preceq [A_{i_{2}}]\preceq ... \preceq [A_{i_{n}}]
\end{equation}
Nothing stops this new ordering to invert the previous one in some instances such that observables invisible at some energy and choice of coefficients become visible under
another choice of coefficients. 
It follows that new ``strong-weak'' dualities can be constructed using the method of coefficient groups. Their applicability goes beyond quantum gravity to subjects like
condensed matter or many particle systems. Everything one has to do is to re-quantize the theory using a different coefficient setup and to take into account possible
torsion groups in homology. While theoretically this is possible it remains to be seen if there are practical difficulties. This will be the subject of another article to be 
developed in the near future. 
Another aspect that might be important in this context is the similarity of these problems with the ``hat problems'' discussed in [7]. The main idea is that although it may
look unlikely, there might exist predictors that after a finite set of trials are always capable of assigning the equivalence class of an operator and determine an order of 
occurrence. These predictors however, depend on the availability of the axiom of choice so they are outside of the scope of this paper. However, their existence may suggest that 
exact locality may be dependent of some very particular choices. 
One may also ask if the renormalization prescription is affected by the indiscernability of states induced by choices of (co)homology.
Possible emergence of new ``topological'' Ward identities (i.e. having their origin in some remaining ``invariance'' under change of topology, prescribed by the UCT) may
 have important roles in a possible renormalization of gravity.

\section{Relativity of Symmetry}

Symmetries are of major importance in physics in general and in quantum field theories in particular. They manifest themselves in the quasi-invariance of an action under the 
transformations of a group. The fact that one has quasi-invariance (i.e. invariance up to a total derivative) of the action under a group may be irrelevant classically, however,
it is important in quantum mechanics as it allows the construction of group-invariant quantum equations (Schrodinger-equations when the group is the non-relativistic Galilei 
group for example).
One may notice that the existence of a quantum formulation of the laws of physics is related to the existence of non-trivial (phase) factors (i.e. additive terms in the composition 
rule of the group operation, see [16]) that cannot be reduced to zero for all group elements (i.e. they form non-trivial classes in the second cohomology of the transformation 
group). One also observes that the existence of basic quantum effects is a result of the global (topological) properties of the groups associated to the supposed ``natural'' 
symmetries
(Galilei group, Lorentz group, conformal group, etc.). These properties are probed via group (co)homologies.
 Information about a group (or in general a space) is not only
encoded in the group (space) itself but also in the way in which the group (space) acts (is mapped) into some reference module (space). This is why one can study group properties
by analysing the actions of the group on an associated space. On that space one can construct a CW complex and analyse it via combinatorial techniques.
Moreover, information about a group (space) may also be encoded in the way in which one probes that group (space).
 One can classify the various ways 
in which information about a group fails to be encoded geometrically (i.e. non-topologically)\footnote{I contrast here geometrical and topological results although they might be
related, see for example Gauss-Bonnet theorem, etc.} by using cohomology groups of different orders. For example the classes of the second 
cohomology group $H^{2}(G,U(1))$ i.e. the cohomology group 
of the maps between the analyzed group $G$ and the unitary $1$-dimensional group $U(1)$ encode the global character of the factors in the composition rule of the group-operation
in $G$ i.e. the way in which they fail to vanish globally [16]. The non-trivial third cohomology group $H^{3}(G,U(1))$ encodes the failure of the associativity property of the
composition rule [16]. 
Also, the existence of not globally vanishing (phase) factors induces super-selection rules.
They are induced in standard quantum mechanics by the presence of non-trivial operators that commute with all
the observables and thus belong to any complete set of commuting observables. As a result, these operators decompose the Hilbert space of all possible states of a system into 
coherent subspaces characterized by their eigenvalues. The superposition principle holds only inside these superselection subspaces and no observable may have non-zero matrix elements
between states of different superselection eigenvalues. As an example one may consider the mass of particles in a space acted upon by a Galilei group. Bargmann superselection rules arising due to 
the topology of the Galilei group forbid for example mass decay (i.e. physical subspaces corresponding to different mass are incoherent). Of course, this is not true as one has to consider the 
Lorentz group as a ``true'' group of nature. What one must remember here is that the existence of such superselection rules is a result of the existence of non-trivial second
group cohomologies of the transformation groups i.e. a result of non-trivial topology of the symmetry group as mapped over a space. 
Further properties can be encoded by higher cohomology groups. However, as showed before, it is important to notice that the topology of a space (or group) cannot be probed in
 an absolute 
sense (regarding all the properties one may wish). In some sense this is an extension of the quantum uncertainty that involves the topology of the space. 
One may quote the existence of super-selection rules in order to avoid solutions like Wheeler's bags of gold. I will show later on that these expectations may be misleading.
In order to
extract useful properties from cohomology one must make a choice of a coefficient structure. Various choices may make classes inside the cohomology merge or become separated. 
The actual ``nature'' of them being ``separated'' or ``merged'' depends on the actual type of ``topological measurement'' (i.e. the choice of a coefficient group). 
Because of this, physical properties depending on classes of (co)homology or being defined as non-trivial function(als) over such classes must have a relative nature. 
As symmetries map various states into equivalence classes one may conclude that symmetries are in general relative.
What I wrote above is visible also in the path-integral formulation. It is well known that anomalies are failures of a symmetry that is manifest at the ``classical'' level i.e. 
in the initial action, to exist after one proceeds to a path-integral quantization. This failure is associated to the non-invariance of the measure of the path integral to the
transformation prescribed by the given group. There are of course physical anomalies (like chiral anomalies) that manifest themselves experimentally and there are gauge anomalies
that must in principle be avoided. In any sense, as seen in [17], relevant anomalies (that cannot be set to zero via ``local'' transformations) are again given by the non-trivial 
BRST cohomology classes at ghost number one on the space of local functionals. They are of course topological in nature and dependent on the way in which the topology of the 
given space (or group) is analysed. In this sense, setting a (global) group structure for the coefficients may prove useful in avoiding gauge anomalies while
 making use of only a limited number of extra dimensions (or none at all). Of course the use of the term ``global'' here may be somehow misleading. These effects are purely 
quantum-gravitational in nature and refer to the situation when the probing of the topology of a space-time region (or a space or group in general)
 becomes uncertain and various choices of coefficient groups in (co)homology become relevant. Please note that this doesn't have to happen only at very high energies or 
low distances. 
\par One should notice that in the case when symmetries are preserved during quantization they are mapped into Ward identities involving Green functions. They have the role of
identifying various Feynman diagrams in the perturbative expansion allowing in this way various proofs of renormalizability for theories that may naively look non-renormalizable
(see Yang-Mills or QCD). One may wonder if suitable splitting of equivalence classes due to various choices of coefficient groups may add supplemental (maybe topological) 
Ward identities that may prove renormalizability of gravity. While this is certainly an interesting subject for meditation it will probably be analysed only in a future paper. 
\section{A conjecture: Relativity of Holography}
Probably the most important result of this paper is the possibility that the Holographic principle is dependent on the choice of the coefficient group. The holographic
 principle states that the non-equivalent degrees of freedom inside a volume can be mapped unambiguously on the surface encapsulating that volume [4].
The key word here is ``non-equivalent''. I proved in theorem 2 that discernability (or equivalence) are relative concepts. 
Following this line of thought the number of non-equivalent degrees of freedom may depend on arbitrary choices. 
In fact one may make a choice of a coefficient group where the number of
 degrees of freedom in a volume largely exceeds the accessible number of degrees of freedom on the encapsulating surface. 
One cannot argue that they are not in the ``observable-super-selection'' sector associated to a measurement because, as showed before, there are situations when there exists a
topological measurement ambiguity (i.e. arbitrary choice of coefficient groups) that makes the existence of such super-selection sectors relative. 
Indeed one may expect that in a complete theory of quantum gravity one cannot count the independent degrees of freedom in the same way as in a classical or non-quantum gravitational theory.
I definitely agree with this. The only difference with respect to the usual interpretation is that there might not be an unequivocal prescription of counting degrees of freedom that is
independent of an arbitrary choice of coefficients. Let me underline that I do not claim that the holographic principle is wrong (or absolutely right by that matter). 
It appears to me that a choice of a coefficient group in (co)homology imposes one form of counting of degrees of freedom (it identifies some as being in the same equivalence class).
It is very likely that for some choices a strict holographic principle emerges. In fact, for a black hole, any group structure that misses the region behind the horizon will
satisfy the standard holographic principle. However, this may not be an absolute property of quantum gravity. I can claim this simply because a general theory of quantum gravity 
should be independent of the choice of coefficients (i.e. topologically covariant) in the same way in which general relativity is diffeomorphism covariant or some 
quantum field theories are gauge invariant. 
Somehow surprising, on the \textit{classical} side there exist solutions of the Einstein field equations that violate the entropy law allowing essentially for an infinite number of degrees of
freedom to be present inside a compact region of space-time. The solutions are called ``Wheeler's bags of gold'' [12]-[13] and are assumed to be eliminated via some 
quantum mechanism mainly in order to obtain results compatible with the AdS/CFT conjecture.
However, it appears to me that the ``bags of gold'' may have some effects after all in a full theory of quantum gravity. They become obvious when one adopts a topological 
definition of entropy in the context presented in this article. 
In order to improve on clarity I start by reminding the standard definition of entropy as being given by the logarithm of the number
 of microstates associated to the same macrostate $S=k_{B} log [\Omega]$ or, when considering a general quantum case the definition becomes 
$S=-k_{B}Tr[\rho Log[\rho]]$ where $\rho$ is the density matrix operator. The entropy can be defined as the failure of macroscopic states to reveal all the microscopic details. 
Otherwise stated it may be interpreted as the uncertainty that remains after a macroscopic state is fully described. 
The concept of entropy evolved from the practical inability of probing classical microstates to the inherent 
inability of probing quantum microstates. An extension would be towards the inability of probing the topological structure of the analysed space and this appears to be
precisely the case when dealing with quantum gravity and coefficient structures in (co)homology.
One may observe that entropy can in general be extracted from the (co)homology of the space of microstates. In fact the cohomology measures precisely
the failure of probing topological structures using local considerations. Because of this, it is a perfect tool for identifying the topological uncertainty i.e. the 
topological component of the entropy. I showed before that this has a measurable effect when a topology is chosen and contributes to the statistics when such a topology 
is left unspecified. Let me call $C$ the space 
of microstates available to a specific microscopic probing of a topological space. This may be represented as a linear combination of simplexes with various coefficients.
 Let $\delta$ be an operator that realizes a form of ``coarse graining'' in the sense of partitioning the microstates into classes according to the macrostates they can 
encode and taking into account the topology of the associated space (i.e. as a boundary operator). Then one can define a chain complex for cohomology as 
\begin{equation}
 ...\xrightarrow{\delta^{n-1}} C_{n-1}^{*}\xrightarrow{\delta^{n}} C_{n}^{*}\xrightarrow{\delta^{n+1}} ...
\end{equation}
or for homology 
\begin{equation}
 ...\xrightarrow{\delta^{n+1}} C^{n}\xrightarrow{\delta^{n}} C^{n-1}\xrightarrow{\delta^{n-1}} ...
\end{equation}
The star in the above description is a notation that makes the difference between homology and cohomology groups manifest. 
The argument here is purely formal. I simply prove that this concept exists. Specific calculations will be the subject of a future article. 
In general the (co)homology group is defined as the group obtained by taking the quotient between the kernel of $\delta^{n}$ and the image of $\delta^{n-1}$. 
In the present context the kernel of $\delta^{n}$ represents the number of microstates that are mapped into the identity class of the space of macroscopic states and 
the image of $\delta^{n-1}$ represents the result of the application of the operator over the initial microstates. 
The (co)homological structure in this case represents the division 
of the kernel in partitions defined by the image. The non-topological entropy may be identified with the number of microstates in a class. Indeed, the class structure is 
not visible macroscopically and contains all the microstates associated to a macrostate. However, this definition offers the advantage of taking into account the additional
topological uncertainty in a more complete way. Different coefficient groups in cohomology may merge or dissociate classes. In this sense entropy is defined only up to a choice
of a coefficient structure over the (co)homology. While the properties of standard entropy remain unchanged if the ``topological uncertainty'' is irrelevant, when this is not
the case (i.e. in the case of strong quantum gravity but not only) entropy can be defined only up to a choice of probing the topology. 
Certain choices of coefficients are known to merge the equivalence classes increasing the total number of equivalent microstates. 
However, each choice of coefficients, once made must remain consistent with itself i.e. no violation of the second law is allowed for any choice. While
a maximum bound may exist for each choice, it may be a relative notion, depending on the actual choice made. 
One must also note that the classification of the topologically distinct features is now encoded in the $Ext$-group (in the case of cohomology) or in the 
$Tor$-group (in the case of homology) via the universal coefficient theorem. The map $Ext(H_{i-1}(X),A)\rightarrow H^{i}(X;Z)\otimes A$ is an injection. 
This means all elements in $Ext$ must have a corresponding element in $H^{i}(X;Z)\otimes A$ but the reverse is not true in general. This means the $Ext$ category 
offers a more accurate classification of ``topologically inequivalent phases'' than would be offered simply from cohomological considerations alone. 
I will not insist on this now but it may prove important in the classification of topological phases. 
As a practical example, I will focus here on the classical solution of Einstein's field equations known as ``Wheeler's bag of gold''. 
In general, the ADM (Arnowitt, Deser, Misner [14]) theory for general relativity allows the foliation of the spacetime manifold into a series of space-like hypersurfaces. The 
next step would be to re-express the Lagrangean in terms of a pure spatial metric ($g_{ij}$), a lapse function $N$ and a shift vector that represents shifts along the tangent 
to the surface of constant time-coordinate. One can now find the conjugate momenta associated to these terms and obtain a Hamiltonian equivalent of the problem. 
In this context solutions to Einstein equations imply the definition of initial data which means the specification of the 3-dimensional Riemanian metric ($g_{ij}$) and its 
conjugate momentum ($\pi^{ij}$). These have to satisfy constraints of the form
\begin{equation}
 ^{(3)}R-(I/g)(\pi^{ij}\pi_{ij}-\frac{1}{2}\pi^{2})=0
\end{equation}
\begin{equation}
 \nabla_{i}\pi^{ij}=0
\end{equation}
where $^{(3)}R$ is the 3-scalar curvature of $g_{ij}$ and $g=det(g_{ij})$ while $\pi^{2}=(Tr\pi^{ij})^{2}$. $\nabla_{i}$ is the covariant derivative corresponding to $g_{ij}$.
Some solutions to these equations possess a ``moment of time symmetry'' i.e. a point where $^{(3)}R=0$. It has been proved [15] that the total energy of an axisymmetric, moment of
time symmetry initial data is positive. One can also write a general expression for an axisymmetric 3-metric of the form
\begin{equation}
 ds^{2}=e^{2q}(d\rho^{2}+dz^{2})+\rho^{2}d\theta^{2}
\end{equation}
However, a metric can be deformed by a conformal transformation of conformal factor $\phi$ leading to another possible solution. 
Suppose now one starts with a smooth conformal factor which is positive at infinity but becomes negative at some point. Obviously it must pass through at least a point where it 
is identical to zero. 
In that point of time all the points on the constant time coordinate surface $S$ are transformed into a single point and must be identified. 
The space becomes the union of an asymptotically flat manifold and a compact manifold. These two are joined at a single point. This solution is called the ``Wheeler bag of gold'' due to the 
singularity appearing at the intersection point. 
In fact one can prove that the energy on one side may become $+\infty$ while on the other side $-\infty$. This formal divergence may be only a classical artifact not to be 
recovered in a full quantum description. However, some relevant quantum effects exist. 
In order to find them one has to integrate over
unequivalent geometric configurations defined by the action 
\begin{equation}
S=\frac{1}{2k}\int R\sqrt{-g} d [vol_{M}]
\end{equation}
where 
\begin{equation}
g=det(g_{\mu\nu})
\end{equation}
$R$ is the Ricci scalar, $g_{\mu\nu}$ is the space-time metric, $k=8\pi G c^{-4}$, $G$ being the gravitational constant, $c$ the speed of light in vacuum and the 
configuration space $\textfrak{E}(M)=(T^{*}M)^{2\otimes}=T^{0}_{2}M$ is a space of rank $(0,2)$ tensors.
It is generally argued that although the classical solutions exist they may be suppressed once the correct measure of integration is used in the quantization prescription.
However, this solution is particularly interesting from the perspective of the universal coefficient theorem. Let me consider a quantum-gravity probing device with an 
internal group structure that can detect the asymptotically flat manifold (say, for example $\mathbb{Z}$). This trivial manifold can be mapped into a ball which
has non-vanishing homology with coefficients in $\mathbb{Z}$ only for the zero dimension. Now attach to this space a sphere $S$ tangent to it at a single point. 
Depending on the group structure used to perform the measurement the sphere may or may not be visible. However, the quantum gravity properties of this structure will 
remain encoded in the possible $Ext$ groups appearing in the UCT sequence. In some sense the information will be encoded in the topology of possible maps of the
group chosen to perform the measurement and the group of the physical spacetime involving a ``bag of gold''. This $Ext$ group is obviously non-trivial
(i.e. the equivalence in standard quantum mechanical language would be ``non-commuting observables''). This requires for the quantization prescription to take the correct
$Ext$ group into account when performing the ``sum over histories''. This allows these types of solutions to indirectly influence the quantum results via
the topologies of the $Ext$ and $Tor$ groups. Of course I do not expect infinite energy in the region covered by the bag of gold as prescribed
in classical general relativity but I also do not expect to have solutions of this type being completely irrelevant in the context of quantum gravity. 
In some sense it is known that processes described by single Feynman diagrams may look non-physical and are certainly unobservable, however, it is the cross section 
calculated with them that makes physical sense. The same situation appears to happen for the bag of gold solutions. While I share the common belief that this solution 
is unlikely to appear as a physical outcome in the sense predicted by classical general relativity (infinite entropy, infinite energy), it appears to me that it should be 
considered in a full theory of quantum gravity simply due to the non-triviality of the extension group it generates. Its overall effect may be the cancellation of some other 
inconsistent object so it might as well never arise as a physical configuration. One could ask if they may 
somehow correlate to the cosmological horizons?

\section{Remarks and applications}
\subsection{Information, measurement and quantum gravity}
As seen in the sections above, the common ideas that appeared to be absolute in the classical (non-quantum-gravitational) approach to physics i.e. observables, symmetries, 
discernibility, entropy, etc. become relative. It is possible that a quantum theory of gravity may not be expressible in terms of local observables and that quantum gravity 
observables must have a rather special form. Analysing the algebraic-topological aspects of gravity it appears that one has to expand the algebraic structures in order
to obtain relevant information. For example in order to probe topologically non-trivial space-times one has to use coefficient groups in cohomology. These may play
the role of an experimental probing device (an apparatus). In this sense an abstract representation of an apparatus in quantum gravity may be seen as a group structure. 
Next, one may ask what procedure has to be performed in order to make a quantum-gravitational measurement. It appears that one has to provide a coefficient group (apparatus)
as an input. The choice of the group structure is not ``predefined'' in the same sense in which the choice of the z-axis in the quantum measurement of a spin 1/2 particle is 
not defined a priori. Once the z axis is defined one may obtain a statistics of the outcomes. In the same sense, once a group structure is defined one obtains a 
(co)homology sequence and an $Ext$ resp. $Tor$ group. The (co)homology obtained in this way will encode the topological properties that can be obtained using the given coefficient
group. The $Ext$ respectively $Tor$ groups will encode the failure of the coefficient groups to encode the full information about the space as well as a means to classify various
choices of coefficient groups i.e. sequences with identical $Ext$ or $Tor$ will form the analogue of symmetry equivalence classes. 
One may also notice that this way of thinking may become useful in the classification of topological phases of matter, apart of the obvious applications to quantum gravity.
One may imagine the quantum gravity measurement device as an extended object that encodes a group structure. The actual measurement is the process of obtaining the
(co)homology (or homotopy) of the given space as an output of the apparatus (i.e. with the coefficient group of the apparatus).
One can regard the UCT as a statement about how much the outcome differs when using an apparatus with a given group
structure with respect to the case when one simply tensors the outcome of an apparatus using a trivial group structure with the previous group structure. This difference is 
encoded in $Tor$ respectively $Ext$ and may be seen as the equivalent of the failure of observables in standard quantum mechanics to commute. 

\subsection{Quantization and topological properties of symmetry groups}
There are several important ideas that come together in this article. On one side I observed that the probing of the topology of a given space or group may be 
fundamentally limited by specific incompatible choices of coefficient structures in the (co)homology. The probing of the topology of a space appears to be limited not only
by a lack of energy or of time as mentioned in some earlier work [11] but also by the fact that certain ``global-measurements'' associated to different coefficient groups in cohomology cannot
 be performed simultaneously in a perfect sense. Some information visible using one choice will be lost when dealing with the other choice.
This fact relativizes certain objects and has various other important effects. The choice of the coefficient structure may determine the topological features that can be
observed. 
In this section I show with some simple examples (following mainly [16]) how some topological 
properties are relevant in the construction of group invariant quantum theories and how quantum effects are actually to be related to the specific behaviour of a theory 
under some symmetry groups. 
In order to keep the discussion as simple as possible I will give the examples using the Galilei group. Its elements can be parametrized by
\begin{equation}
 g=(B,A,V,R)
\end{equation}
where $B$ refers to time, $A$ refers to space, $V$ refers to boosts and $R$ refers to rotations. 
The associated group law is 
\begin{equation}
 g''=g'*g =(B'+B,A'+R'A+V'B, V'+R'V,R'R)
\end{equation}
The action of the group on space-time is obviously 
\begin{equation}
\begin{array}{cc}
 x'=Rx+Vt+A, & t'=t+B
\end{array}
\end{equation}
In classical mechanics one can define a lagrangean as 
\begin{equation}
 L=\frac{1}{2}m\dot{x}^{2}
\end{equation}
This is considered as quasi-invariant as its transformed form differs from the original form only by a total derivative
\begin{equation}
 L\rightarrow L'=L+\frac{d}{dt}m(xV+\frac{1}{2}V^{2}t)=L+\frac{d}{dt}\Delta(t,x;V)
\end{equation}
There is no way of removing the function $\Delta(t,x;g)$ for all transformations $g$ of the Galilei group by adding a total derivative to $L$.
The classical equation of motion (Lagrange equation) is not affected by this change and $\Delta(t,x;g)$ may appear as unimportant although it is relevant when
defining conserved quantities. However, it will reappear in 
the quantum case in an interesting fashion. 
When going to quantum mechanics one identifies the analogue of energy conservation with the Schrodinger equation and in order to keep quantum mechanics Galilei-invariant one must
assure that Schrodinger's equation has the same form in reference frames related via Galilei transformations. 
One may observe that there is no way of implementing Galilei invariance by using a transformation directly on the wavefunction
\begin{equation}
 \psi'(x',t')=\psi(x,t)
\end{equation}
However, one may observe that pure states are in fact described by rays where the set of rays is defined as
\begin{equation}
 \{rays\}=H/R
\end{equation}
where $R$ is the equivalence relation that identifies vectors $\psi$ and $\psi'$ of the Hilbert space $H$ which differ only in an unobservable phase. 
Thus one may enforce Galilei invariance by allowing spacetime dependent phase factors as in 
\begin{equation}
 \psi'(x',t')=exp(\frac{i}{h}\Delta(t,x))\psi(x,t)
\end{equation}
One can determine $\Delta$ by imposing Galilei invariance as
\begin{equation}
 \Delta(t,x)=m(xV+\frac{1}{2}V^{2}t)=\Delta(t,x;g), g\in G
\end{equation}
The exponential is the same as the one appearing in the transformation rule of the Lagrangean. These two functions are caused by related effects. They are in fact related
to the non-trivial cohomology of the Galilei group. 
\par The transformation law given above allows us to find the composition law of two successive transformations
\begin{equation}
 \psi'(x')=[U(g)\psi](gx)=exp(\frac{i}{\hbar}\Delta(x;g))\psi(x)
\end{equation}
where $x'=gx$. If $x''=g'x'=g'gx$ we may write similarly
\begin{equation}
[U(g'g)\psi](x'')=exp(\frac{i}{\hbar}\Delta(x;g'g))\psi(x)
\end{equation}
To compare $U(g'g)$ with $U(g')U(g)$ we first notice that 
\begin{equation}
\begin{array}{c}
 [U(g')U(g)\psi](x'')=[U(g')(U(g)\psi)](g'x')=\\
\\
=exp(\frac{i}{\hbar}\Delta(x';g'))(U(g)\psi)(x')=\\
\\
=exp(\frac{i}{\hbar}\Delta(gx;g'))exp(\frac{i}{\hbar}\Delta(x;g))\psi(x)\\
\\
\end{array}
\end{equation}
Then we obtain 
\begin{equation}
 U(g')U(g)=U(g'g)exp\{\frac{i}{\hbar}(\Delta(gx;g')+\Delta(x;g)-\Delta(x;g'g))\}
\end{equation}
which can be rewritten using 
\begin{equation}
 \xi(g',g)=\Delta(gx;g')+\Delta(x;g)-\Delta(x;g'g)
\end{equation}
as
\begin{equation}
 U(g')U(g)=exp\{\frac{i}{\hbar}\xi(g',g)\}U(g'g)=\omega(g',g)U(g'g)
\end{equation}
where $\omega(g',g)$ are the unimodular factors. This rule defines a projective (or ray) representation of the group G and $\xi$ defines a two-cocycle on G. 
The fact that $\xi$ cannot be made zero for all group elements of the Galilei group (i.e. the projective representation of the Galilei group used in quantum mechanics cannot
be transformed into an ordinary one) is expressed by saying that $\xi$ is a non-trivial cocycle on the Galilei group. 
Since pure states are represented by rays, symmetry operators may be realized by unitary ray operators. These may form equivalence classes bringing together all operators which
differ by a phase that can be locally eliminated. The classes of inequivalent two-cocycles define the second cohomology group $H^{2}(G,U(1))$.
As another interesting example of topological effects on groups is the group extension. The simplest case may be considered the Weyl-Heisenberg group which defines essentially
the quantization prescription. It is a three-dimensional (or in general $(2n+1)$-dimensional) manifold $(q,p,\zeta)$ with the group law given by 
\begin{equation}
\begin{array}{c}
 q''=q'+q\\
\\
 p''=p'+p\\
\\
\zeta''=\zeta'\zeta exp\{\frac{i}{2\hbar}(q'p-p'q)\}\\
\\
(\zeta;q,p)^{-1}=(\zeta^{-1}; -q, -p)\\
\end{array}
\end{equation}
The two-cocyle is here given by 
\begin{equation}
 \xi(g',g)=\frac{1}{2\hbar}(q'p-p'q)
\end{equation}
This two-cocycle is only one representative of its class. One may add two-coboundaries and obtain different but equivalent Lie algebra commutation relations. 
However, preserving the topological structure of the group one cannot globally eliminate these cocycles. One may ask what if the probing of the topological structure
of the transformation group (manifold) may be affected by different choices of coefficients? Would it be possible to merge the identity class with the class of the above cocycle? 
In that case would it be possible to arrive at 't Hooft's conclusion (for example [18]) about ``pre-quantization''? Of course, in this case one must consider possible
Ext-groups for the cohomology exact sequence of the UCT that may return all quantum effects in another way. I will not follow here this line of thought but one must acknowledge G. 't Hooft for his work 
related to this subject albeit he was probably not aware of the algebraic-topological interpretation I present here. I must also underline that the possibility mentioned above is
in essence a quantum effect that merely introduces an ambiguity into the way in which topological properties of groups and spaces can be probed. Standard quantum mechanics
remains valid in each equivalence class. The only difference is that due to further (quantum) uncertainty some equivalence classes may merge when strong gravitational effects 
are present or when special ambiguities in the experimental topological setup are being introduced. 
I also stress that the ``validity'' of quantum mechanics is not altered and this remains a fact, independent of the energy scales, distance scales, etc. 
What I show is only that one may ``abelianize'' the commutation rules of quantum mechanics with the cost of introducing $Tor$ or $Ext$ groups in the chain complex. 
The quantum effects are simply ``shifted'' towards these constructions that must be taken in account in the end of the calculations. 
\subsection{Topology of spacetime and anomalies}
One may ask if my construction is dependent on a purely geometrical interpretation of space-time that may indeed not be valid in the case of quantum gravity. 
In fact there have been several attempts to define quantum-gravity spacetime using a discrete topology (causal sets [19]) or some form of superposition of 
``microscopic geometries'' [20] related to Mathur's ``Fuzzballs'' (essentially fundamental strings that in my representation would be the result of choosing a 
continuous group of coefficients). My approach is a description of why all these approaches are in some sense plausible but still incomplete. 

Considering this, string theory already makes an assumption about the topology of space by introducing the ``worldsheet'' or the ``fundamental string'' in the non-field theoretical 
approach. This might be possible but one has to take into account that by doing this one selects a topology via a group, (say $\mathbb{R}/\mathbb{Z}$ but not necessarily)
 which selects the length of the 
string or the fact that it connects two points. As a consequence string theory can only make predictions for ``experiments'' that are designed in such a way that this configuration 
makes sense. 
Indeed it appears that this offers an UV-completion of the theory and the prediction of the graviton. However, due to its topological non-covariance it must contain an enormous 
amount of irrelevant and/or fictitious information which my idea helps to uncover.
About quantum loop gravity it is known that it introduces a discrete topology of space-time due to its choice of the operator algebra. This too, is an artificial 
construction and focuses the description on ``experiments'' that can probe such a discrete structure. In this case we may speak about the $\mathbb{Z}_n$ group and one has to 
pay attention what fictitious constructions this group generates. Again, the universal coefficient theorem and its exact sequence (with the first injective map) may give an 
image about what dualities one may expect and what objects are non-physical. There is certainly a whole range of alternatives: closed strings, open strings, n-p-branes etc. but 
the reader may
notice that all of them imply choices of topologies hence specific experimental situations that should be probed. They cannot be fundamental for a theory of 
quantum gravity. 

In fact I argue that the topological structure of space-time may be subject to some form of ambiguity in its accurate definition due to the impossibility of probing the
full information encoded in  
topology via (co)homology in an unequivocal way. In this sense the question ``what is the precise topology of space-time at extremely low scales'' may have no precise answer 
unless one provides a specific method of probing that topology. In some sense the problem is similar to the double slit experiment of standard quantum mechanics. 
There, the question ``through what slit did the electron go'' must change the topological setup of the experiment forcing us to obtain a non-interference pattern. If the 
precise trajectory of the electron is of no concern to us the topological setup allows interference patterns. Unlike this case where we can actually control the 
topological setup of the experiment and have a precise definition of it, in quantum gravity this might be fundamentally impossible. One cannot any longer keep all 
topological features independent of the choice of a coefficient structure (i.e. independent of an actual probing of the topology, be it the topology of the space-time itself,
 the topology of the field space or the topological properties of the symmetry groups acting on a given object).
One can notice that anomalies in the construction of a quantum theory of fields may be common and gauge anomalies may appear. This is indeed dangerous for a consistent 
quantum field theory. However, it has been shown that the gauge anomalies are to be associated with classes of the BRST cohomology [17]. Of course, if the topology of the space becomes
uncertain the associated topology of the field space will follow. It can be possible that some choices of group coefficients in (co)homology may make the anomalous cohomology
classes equivalent to the identity (i.e. they become trivial). This doesn't mean that any field theory can be directly quantized but that in the extreme case of quantum gravity 
a choice of coefficients might exist that makes the anomalies cancel in a trivial way. 
I will continue here by analysing the effect on symmetries of the fact that topological properties of groups and spaces depend on choices of coefficient groups in (co)homology. 
Symmetries can in principle be seen as equivalence classes over a space. Different choices of coefficient groups may merge symmetry classes and change the structure of the sets 
of states to be considered equivalent in certain situations. 
One can prove that an anomaly is a loop effect in the Feynamn diagram description. In fact it appears because of the non-invariance of the path integral measure and is encoded 
in the Jacobian of the symmetry transformation. This can be shown to be a loop effect due exclusively to quantization. 
It is well known that one can add in general counter-terms to the classical action as long as they are of higher order in the coupling constant.
 This is because they are corrections to unspecified 
loop terms invisible in the classical theory. This procedure leads to renormalization as long as the added terms are local. 
Let's start with a classical action
\begin{equation}
 S_{cl}=\int d^{4}x (-\frac{1}{4}F_{\mu\nu}^{\alpha}F^{\alpha\mu\nu}+L_{matter}[A,\psi,\bar{\psi}])
\end{equation}
where $\psi$, $\bar{\psi}$ are the matter fields, $A$ is the gauge field and $F_{\mu\nu}$ is the field strength tensor (also for a non-abelian theory).
Suppose there exists a gauge anomaly and suppose one adds a local counter-term of order $3$ in the coupling constant $g$ called $\Delta\Gamma$ such that
\begin{widetext}
\begin{equation}
 S_{cl}\rightarrow S_{cl}+\frac{1}{6}\int d^{4}p d^{4}q \Delta\Gamma^{\mu\nu\rho}_{\alpha\beta\gamma}(-p-q,p,q)A_{\mu}^{\alpha}(-p-q)A_{\nu}^{\beta}(p)A_{\rho}^{\gamma}(q)
\end{equation}
\end{widetext}
At order $g^{3}$ such a term modifies the 3-point vertex function as
\begin{equation}
 \Gamma_{\alpha\beta\gamma}^{\mu\nu\rho}\rightarrow [\Gamma_{\alpha\beta\gamma}^{\mu\nu\rho}]_{new}=\Gamma_{\alpha\beta\gamma}^{\mu\nu\rho}+\Delta \Gamma_{\alpha\beta\gamma}^{\mu\nu\rho}
\end{equation}
If one can find a local $\Delta\Gamma$ such that $(p_{\mu}+q_{\mu})[\Gamma_{\alpha\beta\gamma}^{\mu\nu\rho}]_{new}(-p-q,p,q)=0$ then one says the anomaly is irrelevant.
Whenever such a local counter-term does not exist the anomaly is relevant. One may notice that the ``relevance'' of anomalies is due to their failure to be cancelled locally. 
As stated in the main paper, relevant anomalies can be associated to non-trivial BRST cohomology classes at ghost number one. 
Let now $[\Gamma_{\alpha\beta\gamma}^{\mu\nu\rho}]_{new} \rightarrow [c]$. The arrow maps the transformed 3-point vertex function to a (co)homology class 
of the group $H^{n}(X)$ where $X$ is the associated space. The description here is formal; only the reasoning is of importance. 
Using the UCT one can see that the cohomology group is determined via the short exact sequence:
\begin{widetext}
\begin{equation}
 0\rightarrow Ext(H_{i-1}(X),A)\rightarrow H^{i}(X;Z)\otimes A\xrightarrow{h}H^{i}(X;A)\xrightarrow{r}Hom(H_{i}(X),A)\rightarrow 0
\end{equation}
\end{widetext}

One can now chose $A$ such that the map $X\rightarrow X/([c]\sim id)$ becomes trivial. In this case one cannot distinguish the class of the previously ``relevant'' anomaly from the
identity over $X$. This assures that there exists a coefficient structure over the cohomology that trivializes the anomaly. 
This comes at a cost. One must introduce the extension group on the left $Ext(H_{i-1}(X),A)$. The extension group is generally defined in association with the $Ext$ functor. 
Its definition is not particularly involved: let $R$ be a ring and let $Mod_{R}$ be the category of modules over $R$. Consider $B\in Mod_{R}$, take a fixed $A\in Mod_{R}$ and 
define $T(B)=Hom_{R}(A,B)$ as the set of homomorphisms over $R$ from $A$ to $B$. The $Ext$ functor is defined as
\begin{equation}
 Ext_{R}^{n}(A,B)=(R^{n}T)(B)
\end{equation}
This can easily be calculated considering the injective resolution
\begin{equation}
 0\rightarrow B \rightarrow I^{0} \rightarrow I^{1} \rightarrow ...
\end{equation}
and computing 
\begin{equation}
 0 \rightarrow Hom_{R}(A,I^{0}) \rightarrow Hom_{R}(A,I^{1}) \rightarrow ...
\end{equation}
where we excluded $Hom_{R}(A,B)$ from the complex. Then the extension $(R^{n}T)(B)$ is the homology of this complex. 
So, in the particular case above, the existence of anomalies is ``shifted'' into the way in which one can non-trivially map a general group into an abelian group. 
The relevant information is in this case encoded not in one of the two groups but in the topology of the maps between them. This facilitates calculations for field theories
quantized over cohomologies with particular coefficient groups while preserving the non-trivial information related to quantization in the $Ext$ part of the sequence above.
One should notice that the second arrow in the UCT formula above is an injection i.e. while all the elements of the $Ext$ group must have a correspondence in $H^{i}(X;Z)\otimes A$, the latter group might
have different elements with no correspondence in $Ext$. This may suggest that $Ext$ may be a better measure for the true (physical) anomalies. Indeed, in the standard model
gauge anomalies introduced by chiral fermions cancel naturally when all the fermions are included. However, there appears to be a more general rule suggesting a more accurate method of predicting 
``true'' particles while avoiding to fall in the trap of considering fictitious objects, ``needed'' in order to cancel anomalies, as ``physical particles''.

\subsection{Beyond the Holographic principle}
Finally one may ask what this idea brings new with respect to the interpretation of the holographic principle. In order to answer this I may turn again to the idea of
performing a quantum-gravity experiment. Assume one has a topological measuring device using a particular choice of a group structure for the coefficients. 
It remains to be seen how such a device can be implemented practically. Assume also one performs the measurements at a scale where quantum gravity is irrelevant and 
in a region where there are no black holes to talk of. In this case the choice of the coefficient group is irrelevant. The extension and torsion is always trivial and one
obtains the same results known from simple quantum mechanics. One can chose a complete set of commuting observables and start making predictions considering also the effects
of possible non-commuting observables as it is the custom in standard quantum mechanics.
Now consider a different region of space-time where either because one excites gravitational modes that can alter the topology of spacetime or because one has a black hole 
somewhere, the topology of the space-time stops being trivial. In this case one has to perform a topological measurement with an apparatus that will provide information about 
how the (co)homology or homotopy of the region looks like when seen through the specific choice of the coefficient group. According to this measurement one has to design 
restrictions on the observables allowed by classical quantum mechanics. The $Ext$ and $Tor$ parts of the chain will not be trivial and will have to be considered when 
designing further lower-scale experiments using the space-time measured via the coefficient groups. Not all observables will exist in this situation 
(due to merging of equivalence classes). 
A somehow metaphorical way of looking at this is considering the group choice as a choice of coefficients in a polynomial. Classical quantum measurements after a choice is
 made are 
metaphorically equivalent to finding solutions of these equations. If one chooses for example rational coefficients, the number $e$ (the basis of the natural logarithm) will be 
transcendental (i.e. no polynomial with rational coefficients can have $e$ as a root).

\subsection{Experimental verification}
The idea of adding uncertainty to the topology of space-time itself has, as I showed before, many implications. Unfortunately most of these are not easily verifiable. 
While this article is fundamentally theoretical I try here to pinpoint some possible experiments where this subject may become useful. 
It is known that topology is not only associated to space-time itself. As I showed before and in the main article, one may probe via (co)homology or homotopy with coefficients
(of course in an abstract sense) also field-spaces, groups and other abstract spaces. A more accessible experiment where topological features are important is the Bohm-Aharonov experiment. There, one may observe
the effects of a non-trivial topology generated by a magnetic field, in a region where the given magnetic field vanishes. If one could manage to create a magnetic field
in a state of quantum superposition between a situation with trivial topology and one with non-trivial topology one could check if the measurement of the shift of the 
observed interference pattern will fix the degrees of freedom of the system or if new quantum restrictions may appear due to the quantization of the topology itself. 
One should notice that the topological superposition should ideally be obtained without an entanglement with a local object (like the spin of an electron, etc).
Also, possible verifications could be provided by the study of the topological phases of matter. I expect the procedure given by the UCT to be particularly important for the 
classifications of these phases and for the possible discovery of new ones. The fractional quantum hall effect may also have an interpretation in terms of rational $Ext$ groups. 
One may ask what happens with the theoretical prediction of magnetic monopoles in the context of uncertain topology. Are they still possible?
If future experiments will succeed in proving the fundamental limitations of topological measurements one can safely extend this principle towards space-time itself.

\section{conclusion}
As a conclusion, in this paper I show an aspect of quantization that has been probably overlooked but that may have major implications not only in the description of quantum gravity 
but also in the theory of quantum information. On the quantum information side problems like the ``hat problems'' may have some interesting quantum representations. 
Also possible new ``strong-weak'' dualities may result to be important in fields like condensed matter or many particle physics. 
The discussion of other possible applications in quantum gravity or condensed matter physics will be the main subject of a future article. 

\section{Acknowledgement}
This work is supported by ERC Advanced Investigator Project 267219.
I wish to thank Prof. J. Tennyson for his support during this research.


\begin{thebibliography}{9}
 \bibitem{1}
T. Y. Cao, Cambridge University Press, ISBN:0521634202 (2004)
 \bibitem{2}
C. S. Unnikrishnan, Mod. Phys. Lett. A, 17, 1081 (2002)
\bibitem{3}
C. Adami and G. V. Steeg 2014 Class. Quantum Grav. 31 075015 (2014)
\bibitem{4}
L. Susskind, J. Math. Phys. 36, 6377 (1995)
\bibitem{5}
R. Loll 2008 Class. Quantum Grav. 25 114006 
\bibitem{6}
 B. Dittrich, F. C. Eckert, M. Benito, New J. Phys. 14 (2012) 03500
\bibitem{7}
C. Hardin, A. D. Taylor, The mathematics of coordinated inference: a study of generalized hat problems, Series: Developments in Mathematics, Vol. 33, Springer Verlag (2013)
\bibitem{8}
R. Brunetti, K. Fredenhagen, M. Kohler, Commun. Math. Phys. 180 (1996) 633
\bibitem{9}
A. A. Kirillov, Dynamical systems – 4, Itogi Nauki i Tekhniki. Ser. Sovrem. Probl. Mat. Fund. Napr., 4, VINITI, Moscow, 1985, 141–176 
\bibitem{10}
A. Hatcher, Algebraic Topology, Cambridge University Press (2002) (for the example see Section 2.2 ``Homology with Coefficients'', Example 2.51, page 155)
(for the Universal coefficient theorem see Section 3.A for the Homology case or page 195 for the cohomology case)
\bibitem{11}
N. Lashkari, J. Simon, arxiv 1402.4829v1
\bibitem{12}
J. W. York Jr, J. Math. Phys 14, 456 (1973)
\bibitem{13}
N. O. Murchadha, Class. Quant. Grav. 4 (1987) 1609-1622
\bibitem{14}
R. Arnowitt, S. Deser, C. W. Misner, "Gravitation: an introduction to current research", Louis Witten ed. (Wiley 1962), chapter 7, pp 227--265
\bibitem{15}
D. Brill, Ann. Phys., NY 7 466 (1959)
\bibitem{16}
J. A. de Azcarraga, J. M. Izquierdo, ``Lie groups, Lie algebras, cohomology and some applications in physics'' Cambridge monographs on mathematical physics (1998).
For the non-trivial factors in the composition laws see ch. 3.3, pag. 163.
For the role of the second cohomology group see pag. 161. 
For the role of the third cohomology group and associativity see pag. 186. 
\bibitem{17}
A. Bilal, Lectures on Anomalies, arxiv: 0802.0634v1 (Amsterdam-Brussels-Paris lectures in theoretical high energy physics) (2008)
see pag. 72
\bibitem{18}
G. ’t Hooft, ITP-UU-12/25; SPIN-12/23, arXiv:1207.3612v2
\bibitem{19}
L. Bombelli, J. Lee, D. Meyer, R.D. Sorkin, Phys. Rev. Lett. 59 pag. 521 (1987) 
\bibitem{20}
Samir D. Mathur, arxiv:1401.4097
\bibitem{21}
E. Witten, Phys.Rev. D46 (1992) 5467
\bibitem{22}
R. P. Feynman, Rev. Mod. Phys. Vol 20, No. 2, 1948
\bibitem{23}
S. Merkulov, Class. Quant. Grav. 9 2267 (1992)
\bibitem{24}
Y. Yua, H. Y. Guoa, Phys. Lett. B, vol. 216, 1-2 pag. 68-74 (1998)
\end{thebibliography}
\end{document}